\documentclass[%
 aip,
 numerical,
 rsi,
amsmath,
amssymb,
preprint,%
]{revtex4-1}

\usepackage{comment} 
\usepackage{graphicx}
\usepackage{dcolumn}
\usepackage{bm}
\usepackage[mathlines]{lineno}
\usepackage{hyperref}
\usepackage[utf8]{inputenc}
\usepackage[T1]{fontenc}
\usepackage{mathptmx}
\usepackage{etoolbox}
\usepackage{multirow}
\usepackage{tabularx}
\usepackage{float}

\newcommand{\lithiumCrystalMath}[0]{$\mathrm{LiNbO_3}$}

\newcommand*{\citen}[1]{%
  \begingroup
    \romannumeral-`\x 
    \setcitestyle{numbers}%
    \cite{#1}%
  \endgroup   
}

\newcommand*{\EE}[1]{~\times~10^{#1}}

\makeatletter
\def\@email#1#2{%
 \endgroup
 \patchcmd{\titleblock@produce}
  {\frontmatter@RRAPformat}
  {\frontmatter@RRAPformat{\produce@RRAP{*#1\href{mailto:#2}{#2}}}\frontmatter@RRAPformat}
  {}{}
}%
\makeatother
\begin{document}

\preprint{AIP/123-QED}

\title[Cryogenic Sapphire Oscillator]{Cryogenic sapphire optical reference cavity with crystalline coatings at $1\times 10^{-16}$ fractional instability}

\author{Jose Valencia}
 \homepage{jlvalencia@outlook.com}
\affiliation{
National Institute of Standards and Technology, Boulder, CO, 80302, USA
}

\affiliation{
Department of Physics, University of Colorado, Boulder, CO, 80309, USA}%

\author{George Iskander}
\affiliation{Department of Physics, University of Chicago, Chicago, IL, 60637, USA}
\affiliation{Argonne National Laboratory,  Lemont, IL, 60439 USA}

\author{Nicholas V. Nardelli}
\affiliation{
National Institute of Standards and Technology, Boulder, CO, 80302, USA
}
\affiliation{
Department of Electrical Engineering, University of Colorado, Boulder, CO, 80309, USA}%

\author{David R. Leibrandt}
\altaffiliation{Present address: Department of Physics and Astronomy, University of California, Los Angeles, CA}
\affiliation{
National Institute of Standards and Technology, Boulder, CO, 80302, USA
}

\author{David B. Hume}
 \homepage{dbh@nist.gov}
\affiliation{
National Institute of Standards and Technology, Boulder, CO, 80302, USA
}%
\date{\today}

\begin{abstract}
The frequency stability of a laser locked to an optical reference cavity is fundamentally limited by thermal noise in the cavity length.  These fluctuations are linked to material dissipation, which depends both on the temperature of the optical components and the material properties. Here, the design and experimental characterization of a sapphire optical cavity operated at 10 K with crystalline coatings at 1069~nm is presented. Theoretical estimates of the thermo-mechanical noise indicate a thermal noise floor below $4.5\times10^{-18}$. Major technical noise contributions including vibrations, temperature fluctuations, and residual amplitude modulation are characterized in detail. The short-term performance is measured via a three-cornered hat analysis with two other cavity-stabilized lasers, yielding a noise floor of $1\times10^{-16}$.  The long-term performance is measured against an optical lattice clock, indicating cavity stability at the level of $2\times10^{-15}$ for averaging times up to 10,000 s.
\end{abstract}


\maketitle

\section{\label{sec:level1}Introduction to optical reference cavities}

Ultrastable optical reference cavities are crucial tools for tests of fundamental physics~\cite{braxmaier2001, muller2003, muller2003_2, wiens2016, wcislo2018, geraci2019, kennedy2020}, gravitational wave detectors~\cite{kwee2012}, and precision frequency measurements with optical atomic clocks~\cite{ludlow2015}.  Cavities based on high-finesse mirrors optically contacted to monolithic spacers have demonstrated the highest levels of stability. To minimize the effect of temperature fluctuations, room temperature cavity spacers are typically made of ultra-low expansion (ULE) glass~\cite{young1999, jiang2011, martin2013, hafner2015, schioppo2022}, whereas cavities operated at cryogenic temperatures often use crystalline materials (silicon or sapphire), which exhibit low coefficient of thermal expansion (CTE) at low temperatures~\cite{taylor1995, seel1997, matei2017, kedar2022, wiens2023}. The highest stability cavities are currently based on spacers and mirror substrates made of single-crystalline silicon at a temperature of 123 K~\cite{matei2017}, where silicon has a zero-crossing of its CTE, reaching a fractional frequency instability as low as $\mathrm{4\times10^{-17}}$.  Optical cavities with even higher stability would allow for more stable optical clocks and more stringent constraints of fundamental physics.

The highest stability cavities reach a fundamental limit based on thermal noise related to mechanical dissipation in cavity components~\cite{levin1998, braginsky1999, braginsky2003}. The amplitude of thermal noise depends mainly on material choices and the temperature of the cavity components, where stiffer materials with low mechanical loss and lower temperatures improve the thermal noise limit. The leading source of dissipation in most ultrastable optical cavities is the high-finesse optical coatings that define the mirror surfaces~\cite{kessler2012Thermal}. This limitation could be improved using a newer coating technology, like GaAs/AlGaAs crystalline coatings, which exhibit a lower loss angle~\cite{cole2013}.  

To reach the thermal noise limit, numerous sources of technical noise must be carefully suppressed.  These arise from perturbations that affect the length of the cavity directly and those that affect the precision with which a laser can be locked to the cavity.  Common sources of cavity length fluctuations are vibrations, which cause elastic deformation of the cavity, and temperature drifts, which cause thermal expansion and contraction.  A common source of perturbations to the laser lock is residual amplitude modulation (RAM), which can vary in time and cause offsets to the Pound-Drever-Hall error signal typically used for laser feedback.   




Here, an ultrastable cryogenic sapphire optical cavity is described, which is designed to achieve low thermal noise and to limit technical sources of noise, like vibrations and temperature drifts. It uses crystalline GaAs/AlGaAs mirror coatings with a thermal noise limit estimated to be below $\mathrm{4.5\EE{-18}}$ at the operating temperature of 10 K. Operating at this temperature not only lowers the thermal noise limit but also reduces the coefficient of thermal expansion, suppresses heat transfer due to blackbody radiation, and improves vacuum pressure.  Sapphire is chosen because of its high elastic modulus, low mechanical loss, transparency at a laser wavelength of 1~$\rm \mu m$ and its availability in large single-crystal boules \cite{dobrovinskaya2009}.  A closed-cycle cryostat is used and is mechanically isolated from the cavity to achieve continuous operation while limiting vibrations transferred to the cavity.  Several layers of heat shields reduce temperature deviations that affect the long-term cavity stability. This work demonstrates the control of these noise sources at a level sufficient to achieve instability of $\mathrm{1\EE{-16}}$.

The paper is organized as follows. Section II  details the objective of the cavity design within the scope of thermal noise, acceleration sensitivity, cryostat design, and vacuum design. Section III characterizes sources of noise due to residual amplitude modulations, vibrations, and temperature fluctuations. Section IV reports the cavity stability measured with three-corner hat for short averaging times and frequency comparison against an optical lattice clock for long averaging times. Section V summarizes the results and discusses future improvements to the system.

\section{Cavity and apparatus design}

\subsection{Cavity Design}

The main objective of this optical cavity is to stabilize a laser local oscillator for an optical clock based on the $\mathrm{^1S_0 \rightarrow  {}^3P_0}$ clock transition in $\mathrm{{}^{27}Al^{+}}$ \cite{brewer2019}. The wavelength of this transition is 267.4 nm (frequency, ${\nu_0 = 1.121\times10^{15}}$~Hz) and the excited-state lifetime is 20.6 s (natural linewidth, ${\Delta\nu_0 = 7.7}$~mHz)~\cite{rosenband2007}.  The relative linewidth of ${\Delta\nu_0/\nu_0 = 6.9\times10^{-18}}$ sets an approximate target on the laser instability required to interrogate the natural linewidth of the transition.  The relevant averaging times for this level of stability are related to the probe time and the time constant for locking the laser to the atomic transition.  Lifetime-limited stability is reached for probe times near the excited-state lifetime with a servo time constant typically a factor of 10 longer than the probe time~\cite{beloy2021}.  With these considerations in mind, cavity stability is most relevant for timescales between 1 s and 100 s.  Simulations of the clock servo indicate that the ultimate quantum limit on stability imposed by the excited state lifetime for a single Al$^+$ ion, $\sigma_y(\tau)\sim6\times10^{-17}/(\tau/{\rm s})^{1/2}$, requires a thermal noise floor of $3\times10^{-18}$. Similar considerations apply to other clock species, where lifetimes of the clock transition can be longer and noise due to the Dick effect can be a limitation \cite{ludlow2015}. 

The next two sections address design choices for the optical cavity aimed at reaching those limits.  First, thermal noise (Sec.~\ref{sec:thermal_noise}) sets a fundamental limit to the cavity instability based on material properties of the cavity spacer, mirror substrates and mirror coatings, which will be observed when all technical sources of noise are reduced below that limit.  Second, the sensitivity of the cavity to accelerations applied to the cavity mount are determined by the same material and geometric design choices (Sec.~\ref{sec:acc_sens}).

\subsubsection{Thermal noise\label{sec:thermal_noise}}
Thermomechanical noise is caused by thermally-driven fluctuations, which affect the distance between the mirrors' surfaces.  This inherent instability of the cavity length depends on the cavity temperature and material properties like the modulus of elasticity and mechanical loss tangent~\cite{callen1951, levin1998, kessler2012}, $\mathrm{tan(\phi_{loss~angle})}=1/Q$, where $\mathrm{Q}$ is the quality factor.  Two complementary approaches to reduce the thermal noise limit for optical cavities have been pursued.  The first reduces the effect of thermal noise by scaling systems up to longer cavity lengths \cite{} and larger laser beam waists on the mirror surfaces \cite{}, which spatially averages over thermal noise fluctuations.  The second approach is to reduce the magnitude of the thermally-driven fluctuations by cryogenically cooling the cavity and using crystalline materials with improved mechanical properties for the cavity components.  To reach the design goal, a sapphire spacer of length $L=20$~cm and sapphire mirror substrates with 1 m radius of curvature were chosen, with an operating temperature of $T=10$~K.  The dominant source of thermal noise in optical cavities is typically related to loss in the mirror coatings themselves~\cite{kessler2012}.  To reduce this source of noise GaAs/AlGaAs crystalline coatings were applied via optical bonding to the sapphire substrate~\cite{cole2013}.  In what follows, we review results from theoretical investigations of thermal noise and show that the analytical estimates are consistent with our design goal of $3\times10^{-18}$.        

Previous work \cite{levin1998} has applied the fluctuation-dissipation theorem\cite{callen1951} to relate the amount of dissipated thermal energy in the optical components to cavity length fluctuations.  Specifically, the power spectral density of cavity length fluctuations at temperature $T$ can be related to the thermal energy, $W$, dissipated by a fictitious sinusoidal driving force with amplitude $F_0$ at frequency $f$ by
\begin{equation}
\label{eqn:thermalNoisePSD}
S_x(f)=\frac{4k_BT}{\pi^2 f^2}\frac{W_{spacer}+W_{mirror}}{F_0^2}.
\end{equation}
Here $k_B$ is the Boltzmann constant and the thermal energy has been separated into components corresponding to the cavity spacer and the cavity mirrors.  The contribution from both the mirror substrate and the mirror coating have been included in a single term because they follow from a similar analytical treatment~\cite{harry2002}.  

The contribution from the cavity spacer has been estimated by considering the cavity as a hollow cylinder with a length of $L$, an outer diameter of $R$, and an inner diameter of $r$ using~\cite{kessler2012Thermal}
\begin{equation}
\label{eqn:Uspacer}
\frac{W_{spacer}}{F_0^2} = \frac{f L}{(R^2-r^2)}\frac{\phi_{spacer}}{E_{spacer}}, 
\end{equation}
where $E_{spacer}$ is the material elastic modulus and $\phi_{spacer}$ is its mechanical loss tangent.  Similarly, the contribution from the mirrors can be estimated by assuming the size of the laser waist at the mirror surface, $w_0$, is much less than the inner diameter of the spacer such that the mirror can be treated as an infinite half space.  The thermal noise will, in general, involve anisotropic loss angles from the crystalline materials that make up the substrates and coating layers, which are not well-characterized experimentally.  For our estimates, we assume isotropic loss for all materials.  Anisotropic deformation of the the materials also affects the values of the predicted noise floor of the cavity, but assuming a small Poisson ratio, $\sigma\ll1$, for all materials leads to errors below $\sim20$~$\%$.  Finally, with sapphire chosen as the substrate material, the approximation $E_{coating} \ll E_{substrate}$, allows for a simplified expression:
\begin{equation}
\label{eq:Usubstrate}
\frac{W_{mirror}}{F_0^2} = \frac{2\sqrt{\pi} f}{w_0}\left(\frac{\phi_{substrate}}{E_{substrate}} + \frac{\phi^{\prime}_{coating}}{E_{coating}}\right).
\end{equation}
The effective loss angle of the mirror coating is defined here as 
\begin{equation}
\label{eq:2substrate}
\phi^{\prime}_{coating} \equiv \frac{1}{\sqrt{\pi}}\frac{d}{w_0}\phi_{coating},
\end{equation}
where $d$ is the thickness of the coating and $\phi_{coating}$ is its measured loss angle~\cite{cole2013}. All relevant parameters for these thermal noise estimates are summarized in Table~\ref{tab:thermal_noise_params}. 

\begin{table}
\begin{center}
\begin{tabular}{ | m{2cm} | m{2cm} | } 
 \hline
 Parameter & Value \\
 \hline
 $E_{spacer}$, $E_{substrate} $ & 345~GPa~[\citen{dobrovinskaya2009}] \\
\hline
 $E_{coating}$  & 100 GPa~[\citen{cole2013}] \\
\hline
 $\phi_{spacer}$, $\phi_{substrate}$ & $1\times10^{-8}$~[\citen{uchiyama1999}] \\
\hline
 $\phi_{coating}$ & $4.5\times10^{-6}$~[\citen{cole2012}] \\
\hline
 $L$ & 20 cm \\
\hline
 $R$ & 14 cm \\
\hline
 $r$ & 4 cm \\
\hline
 $w_0$ & 337~$\rm{\mu m}$ \\
\hline
 $d$ & 6.37~$\rm{\mu m}$ \\
\hline

\end{tabular}
\end{center}
\caption{\label{tab:thermal_noise_params}Material properties and cavity geometric parameters relevant for thermal noise calculations. The dielectric loss tangent of sapphire is $\mathrm{\sim100}$ times lower than the crystalline coating loss angle, so the spacer and substrate thermal noise contributions are negligible.}
\end{table}

The power spectral density of noise for both the spacer and mirror terms scale with frequency as $1/f$, corresponding to a flat noise floor in the frequency stability of the cavity when characterized by the Allan deviation~\cite{riley2008}.  The fractional frequency instability of this noise floor is given by
\begin{equation}
\label{eq:thermalAvar}
\sigma_{y} = \sqrt{2\ln{2}\frac{S_x(f) f}{L^2}}.
\end{equation}
For the parameters and approximate expressions given above, the thermal noise floor of this cavity is expected to be ${\sigma_{y} = 2\times10^{-18}}$, with the coating contribution accounting for 91~\% of the noise power, and the cavity spacer contributing a negligible fraction ($<0.1~\%$).  Using more complete expressions for the coating thermal noise from Ref.~\cite{harry2002} does not change these estimates significantly.  The largest uncertainty likely comes from the value of $\phi_{coating}$, which has not been studied extensively at cryogenic temperatures.  Taking, instead, the better established room-temperature value of $\phi_{coating}=2.5\times10^{-5}$\cite{cole2013}, a conservative estimate for the cavity thermomechanical noise floor is $\sigma_y = 4.5\times10^{-18}$.

Other sources of thermal noise have also been considered in the literature~\cite{harry2012}; these include temperature-driven fluctuations in the thermal expansion and index of refraction of the substrate and coating materials.  The sensitivity of the mirror materials to these effects reduces significantly at low temperature.  In addition, relative to thermo-mechanical noise with a $1/f$ noise spectrum, thermo-optic noise tends to be less significant at low frequencies $f<10$~Hz where cavity stability is most important in the present application.  For these reasons, cavity thermo-optic noise is expected to be negligible.     

\begin{figure}[ht]
\includegraphics[width=50mm]{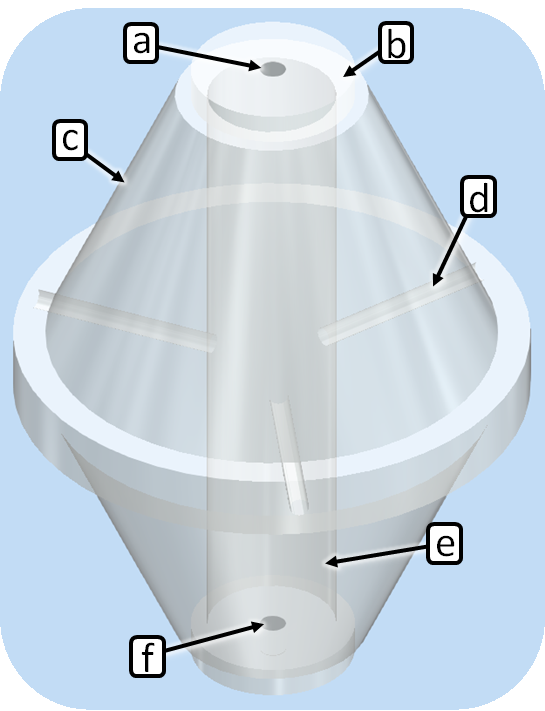}
\caption{\label{fig:cavityDescription} Optical cavity components. The cavity includes a spacer (c) made of single-crystalline sapphire with a central bore hole (e) along the $c$-axis for the optical mode and vent holes (d) along the $a$-axes. The substrates (b) have high-finesse, GaAs/AlGaAs crystalline-coatings on their inner surfaces (f).  The back sides of the substrates (a) are wedged at 0.5 deg. and anti-reflection coated to minimize parasitic etalons.}
\end{figure}

\subsubsection{Acceleration Sensitivity\label{sec:acc_sens}}

Acceleration sensitivity is a measure of how much the cavity length changes as a function of accelerations applied to its mounting points.  It is typically expressed in fractional frequency units divided by acceleration (e.g. in units of $1/g$, where $\mathrm{g=9.8~m/{s^2}}$). The cavity acceleration sensitivity can be modified by engineering the geometry of the cavity and its mount. 

The present design follows that of Refs.~\cite{kessler2012, matei2017} by making the cavity optical axis vertical, along the $c_\parallel$-axis of the spacer and substrates.  The cavity is symmetric about its midplane, consisting of two conical halves separated by a cylindrical shelf of 2 cm thickness and 16 cm outer diameter.  The bore hole along the optical axis is 4 cm in diameter and the mirror substrates are 5.08 cm in diameter and 0.95 cm thick as seen in Fig.~\ref{fig:cavityDescription}.  The mirror substrates also have the optical axis ($c$-axis) along the cavity mode, but the orientation of the $a$-axes of the substrates were not controlled relative to the spacer $a$-axes. Instead, the mirror coating crystalline axes, which have birefringence, were oriented to maximize the splitting between the two polarization modes of the cavity.  The cavity structure exhibits three-fold (120 deg.) rotational symmetry due to the hexagonal structure of the crystalline lattice~\cite{dobrovinskaya2009}. This symmetry is maintained by drilling three vent holes in the cavity along the $a$-axes and mounting the cavity on three conical polytetrafluoroethylene (PTFE) bumpers positioned in an equilateral triangle beneath the central rim, 7.5 cm away from the optical axis.  

Considering the translational degrees of freedom, this cavity structure is nominally insensitive to horizontal accelerations by symmetry. However, imperfections in the cavity geometry could lead to increased horizontal acceleration sensitivities \cite{milo2009}. To address this problem, machining and assembly tolerances were considered in conjunction with design parameters such as the cavity length, cavity diameter, and taper angle\cite{lisheng2006}. The largest effect identified was the misalignment of the c-axis of the cavity spacer with the symmetry axis of the cavity, which was specified at $\mathrm{\pm}$2 deg, leading to approximately $\mathrm{3\EE{-11}}$ uncertainty in the horizontal acceleration sensitivity. By choosing the orientation of the mounting points relative to the cavity vent holes, and hence the $a$-axis of the sapphire crystal, the sensitivity of the cavity to vertical accelerations can be nulled~\cite{matei2016}. Measurements pertaining to the acceleration sensitivity and ambient vibrations are described in Section \ref{sec:vibAcc}.

\begin{figure}[ht]
\centering
\includegraphics[width=85mm]{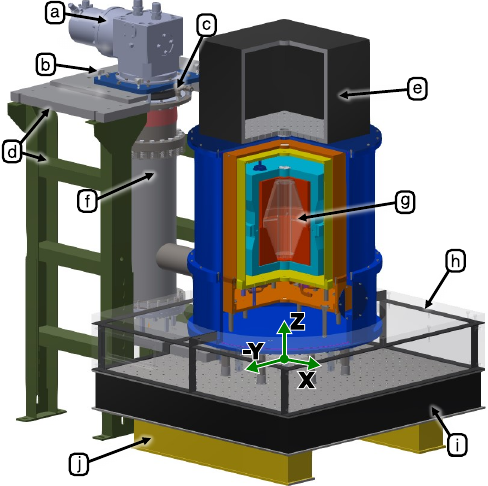}
\caption{\label{fig:wholeApparatus} Experimental apparatus including cryogenic, mechanical and vacuum components. False color is used to differentiate components. The coordinate system (used for acceleration sensitivity measurements) identifies the Z direction along the cavity optical axis and X and Y directions in the horizontal plane as shown. The cryocooler (a) is mounted to an aluminum plate with freedom to be translated and tilted (b), then locked in place to a stand (d) that is bolted to the floor. The vibrations from the closed-cycle cryocooler are decoupled from the experiment through the use of a gaseous exchange space (f), where a rubber bellows (c) contains helium gas that  transfers cooling power to the cavity. The cavity (g) is shielded from temperature fluctuations via multiple heat shields and vacuum chambers that are mounted on an optical table. The optical table (i) for the incident laser is enclosed with an acrylic box (h) and is isolated from vibrations coupled through the ground via active vibration isolation legs (j). Above the cavity is a breadboard (e) containing optics and electronics that monitor the cavity transmission.}
\end{figure}

\subsection{Cryostat Design and Cooldown}
Figure \ref{fig:wholeApparatus} shows the main components of the apparatus including the commercial cryostat~\cite{coldedge}, vacuum chambers, and optical table.  The cryostat consists of a closed-cycle pulse-tube that is held on a separate stand from the optical table to reduce transmitted vibrations. The cryocooler transfers cooling power convectively via a gaseous helium exchange space to a cold stage within the chamber.  By using rubber bellows to seal the helium exchange space and annealed copper braids to transfer heat to the cold stage, vibrations from the cryocooler are reduced at the cavity cryostat chamber. 

\begin{figure}[ht]
\centering
\includegraphics[width=85mm]{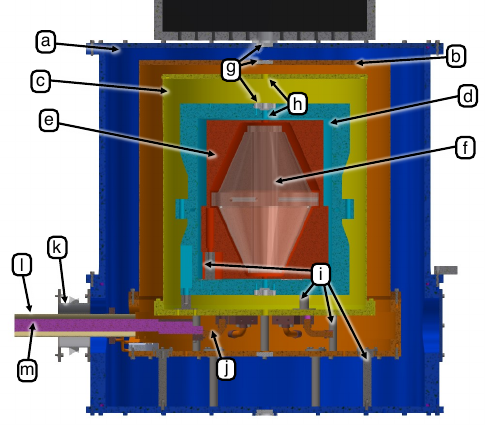}%
\caption{\label{fig:cavityInside} Cryostat assembly. False color is used to differentiate parts of the system. The outer vacuum chamber (a) and heat shield (b) allow optical access through wedged and tilted BK-7 windows (g) at the top and bottom. The heat shield is connected to the helium exchange space through a copper tube (l) and copper braids. The coldplate (c) has a vent hole (h) for optical access and it is connected to the cryostat cold finger (m) through a copper rod and copper braids (j). Steel bellows (k) surround the cold finger and connect to the helium exchange space. The inner vacuum chamber (d) further shields the cavity mount (e) and cavity (f) from radiative heat load and protects the cavity mirrors from contaminants. Each shield is supported by lengthened and hollowed stainless steel posts (i).}
\end{figure}

The outer vacuum chamber, heat shield, cold plate, inner vacuum chamber, cavity mount, and cavity are shown in Fig. \ref{fig:cavityInside}. A base temperature of 8 K for the inner cold plate and 155 K for the outer heat shield is reached. The temperatures of the inner cold plate and outer heat shield are then stabilized with resistive heaters to 10 K and 160 K to minimize temperature fluctuations. The temperature is increased to give headroom of the baseline temperature slowly raising over months of operation.

\begin{figure}[ht]
\centering
\includegraphics[width=85mm]{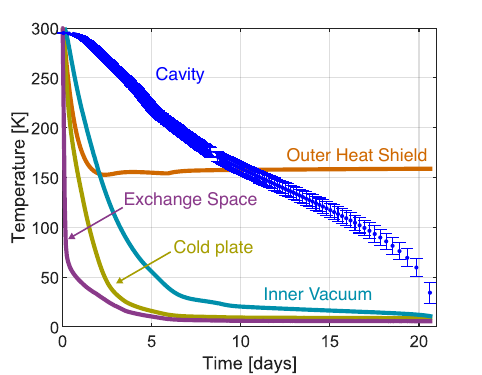}%
\caption{\label{fig:cooldownFinesse} Temperatures of the cryostat components during cooldown. The outer heat shield reaches a baseline of 155 K and is subsequently stabilized to 160 K. The coldplate base temperature is approximately 8 K and is subsequently stabilized near 10 K. While the temperature sampling of the heat shields are nearly-continuous, the cavity is measured with optical transits from a laser and has discrete time steps. There is a gap in optical transits (day 8) when the pre-stabilized laser unlocked, but the drift at that time was nearly linear and the missed transits were interpolated.}
\end{figure}

Figure~\ref{fig:cooldownFinesse} shows the temperature of each heat shield and the cavity during a cooldown. Temperature sensors are located at the exchange space (on the side closest to the cavity), the outer heat shield, the cold plate, and the inner vacuum chamber. There are no temperature sensors located inside the inner vacuum chamber, so the cavity temperature during cooldown is estimated by observing the cavity frequency as a function of time. To do this, a pre-stabilized laser is aligned to the cavity $\mathrm{TEM_{00}}$ mode, and the cavity transmission is measured as a function of time.  Due to thermal contraction, the cavity mode is periodically on resonance with the incoming light, and a "transit" of the cavity resonance is recorded.  An assumption is that the cavity begins the cooldown in thermal equilibrium with the inner vacuum chamber, so the temperature can be estimated based on the known temperature dependent thermal expansion coefficient of sapphire~\cite{gaal1999} and the time stamps of the transits.  The error bars in the cavity temperature are based on a quadrature sum of contributions from the initial temperature uncertainty of the cavity and the uncertainty of the thermal expansion coefficient. 

\begin{figure}[ht]
    \centering
    \includegraphics[width=85mm]{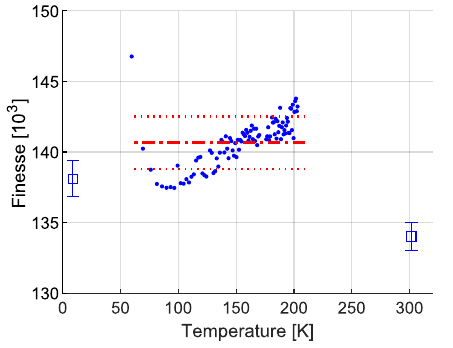}
    \caption{Cavity finesse during cooldown.  The red, dashed line is the mean of the blue circular dots with the red, double-dashed lines showing $\mathrm{\pm\sigma}$ error bars. The blue boxes are finesse ring-down measurements before and after cooling down with $\mathrm{\pm\sigma}$ error bars from the ring-down fit. Early data samples were disregarded because of $\mathrm{\sim}$1 kHz noise in the pre-stabilized laser that affected finesse measurements.}
    \label{fig:FoverT}
\end{figure}

The frequency drift rate of the cavity during a transit can be determined based on its known free-spectral range and the timestamps for each transit.  Using this information, recorded linewidths can also be used to monitor the finesse during cooldown. Fig.~\ref{fig:FoverT} shows the cavity finesse as a function of temperature. Additionally, before and after cooling down the system, transmission ringdown measurements are used to separately determine the finesse. A $\simeq2$~\% increase in the finesse is observed when the cavity is cold based on ringdown measurements before and after the cooldown.  The finesse estimates during cooldown agree with these values within 5\%, but they can be affected by nonlinear terms in the cavity drift rate, which are pronounced at low temperature. 

\subsection{Vacuum Design}
The cavity cryostat has two separated vacuum chambers: the outer vacuum chamber enclosing all cryostat components and the inner vacuum chamber enclosing the cavity and its mount. There is vacuum conductance between all shield layers outside the inner vacuum chamber via small apertures. The inner vacuum chamber is isolated to prevent dust and other contamination on the high-finesse mirrors. Despite this, during the first cooldown of the system, a sudden loss of finesse near the base temperature was observed, which, based on observations after warming up and disassembling the cavity, was found to be due to dust on the lower mirror.  Following cleaning of the mirrors and reassembly of the cavity, loss of finesse has not been observed during subsequent cooldowns.  During each cooldown, the cavity is the warmest component of the system, so cryopumping on colder components of the system prevents residual gases from affecting mirror finesse.    During warmup, where the optical cavity is the coolest object, particles can cryopump on to the mirrors. Loss of finesse from $F=130000$ to $F=100000$ has been observed during a warmup.

The inner vacuum chamber is comprised of two aluminum halves sealed with a spring-energized metal o-ring. Windows (tilted and wedged at 0.5 deg to prevent etalons) are positioned along the optical axis at the top and bottom of the chamber and sealed using indium o-rings. In addition, there is a stainless steel adapter mated to the top of the chamber using an indium seal that allows for a copper pinch-off tube for pumping out the chamber.  After five cooldowns and warm ups, neither indium nor metal o-ring seals need to be replaced unless purposefully opened.  Before assembling the heat shields, the inner vacuum chamber is pumped with a turbomolecular pump to $\mathrm{\sim10^{-9}~mbar}$ (1 mbar = 100 Pa) and is then pinched off.  Maintaining low vacuum pressure inside the cavity relies on cryopumping on the surfaces of the inner vacuum chamber components. 


The outer vacuum chamber is sealed with a viton o-ring. It is pumped out with a turbomolecular pump to $\sim10^{-5}$~mbar before cooldown.  When cooled to 10 K, the outer vacuum chamber maintains a pressure below $\mathrm{3\times10^{-7}~mbar}$. A 10 L/s noble diode ion pump on the outer vacuum chamber pumps residual hydrogen and helium that may not be cryopumped.

\section{Characterization of noise sources}

\begin{figure}[ht]
\centering
\includegraphics[width=85mm]{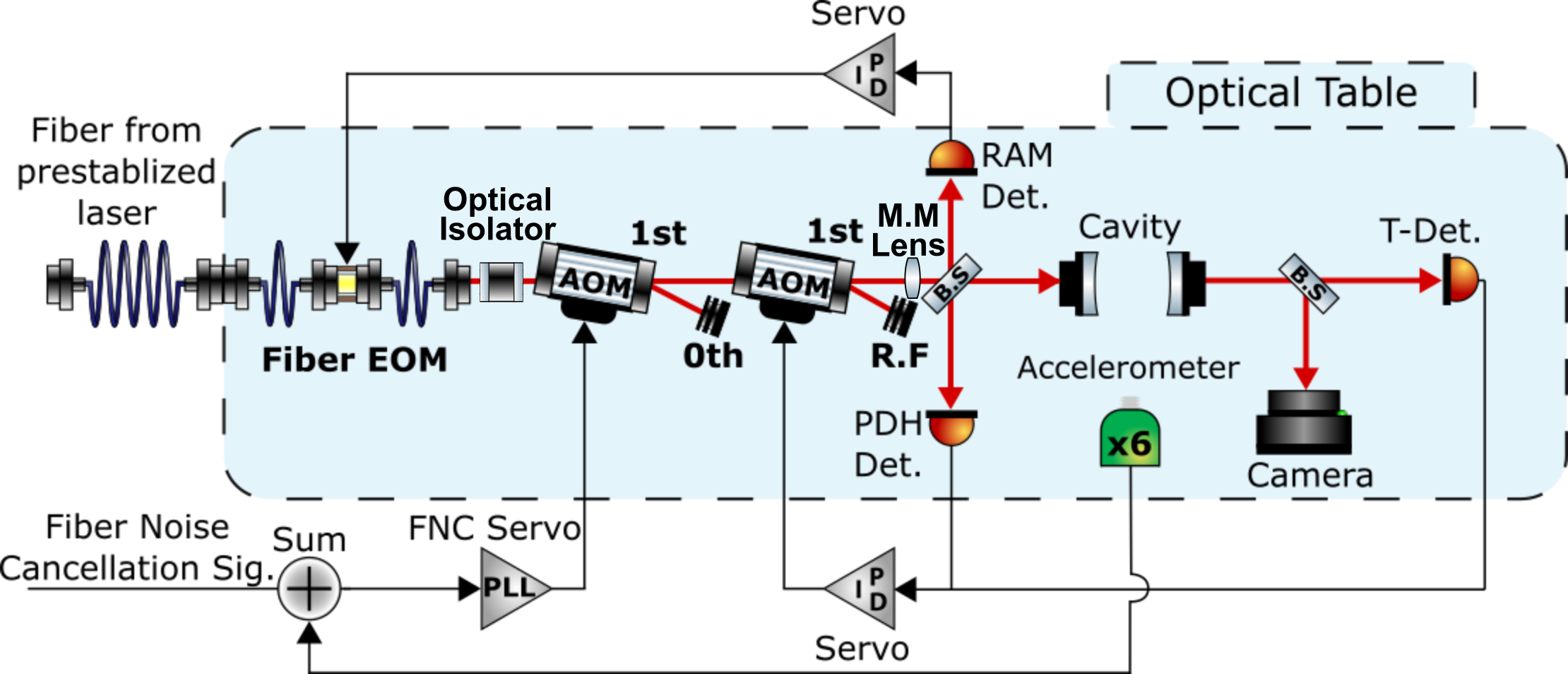}%
\caption{\label{fig:oscarPDHsetup} Optical elements used for laser stabilization and characterization of the cavity.  Pre-stabilized laser light at 1069~nm is delivered via an optical fiber. The light is phase modulated with a fiber-coupled waveguide EOM. The light is collimated and  passes through an optical isolator and through two AOMs. The first AOM is used to compensate for laser phase noise caused by the optical fiber and the second AOM is used to lock the laser frequency to the optical cavity and stabilize the laser power transmitted through the cavity. After the second AOM, the non-diffracted light is retroflected to create a beatnote for fiber-noise stabilization. A mode-matching lens is used to mode match the laser to the $\mathrm{TEM_{00}}$ mode of the optical cavity.}
\end{figure}

A schematic diagram of the most relevant optical and electronic components of the cavity system is shown in Fig.~\ref{fig:oscarPDHsetup}. The laser light incident on the cryogenic cavity is prestabilized to a room-temperature ULE cavity~\cite{young1999}. The light is delivered via optical fiber to a fiber coupled waveguide electro-optic modulator (EOM), and then it is collimated in free space towards the cryogenic cavity. Two free-space acoustic optical modulators (AOMs) are used, one to cancel phase noise in the fiber, and one to lock the laser frequency to the cryogenic cavity resonance. Separate photodetectors are used to detect the Pound-Drever Hall (PDH) error signal, monitor residual amplitude modulation (RAM), and to servo the optical power transmitted through the cavity. Six accelerometers, which are rigidly mounted to the optical table, measure accelerations in all rigid-body degrees of freedom. 

\subsection{Residual Amplitude Modulation}
For PDH locks, a common source of technical noise is residual amplitude modulation (RAM) at the PDH phase modulation frequency, quantified in terms of its modulation depth. This leads to offsets in the error signal that can drift in time, which is written onto the laser frequency through the servo. There are two primary physical mechanisms that create RAM. The first is misalignment of the polarization of the incoming light with the principle axes of the EOM crystal, such that the interaction between light and the crystal does not only  modulate the phase but also modulates the polarization. Any polarizing element after the EOM will convert the polarization modulation to amplitude modulation. The second mechanism is parasitic \'etalons in the optical path, which lead to offsets in the PDH error signal that can vary due to vibrations, temperature drifts and other environmental effects.  To combat these effects, previous studies have used Brewster-cut crystals \cite{bi2019}, temperature-stabilized EOMs \cite{shen2015}, and voltage-stabilized waveguide EOMs\cite{Zhang2014} to reach RAM ratios as low as 1 ppm.


Here, the EOM in use is a waveguide lithium niobate (\lithiumCrystalMath) crystal. \lithiumCrystalMath~is known to have an index of refraction dependent on temperature and applied DC voltage, so it is possible to minimize RAM by servoing the temperature and the DC voltage of the crystal. The smaller volume of a waveguide EOM crystal enables high servo bandwidth compared to free space geometries. The temperature-dependent index of refraction is measured by changing the temperature in steps and measuring the range of the RAM output signal while sweeping twice the half-wave voltage $\mathrm{V_{pi}}$. The output range of the voltage signal maps the amplitude dependence as a function of temperature steps. A periodic temperature dependence is observed with a period of $\mathrm{\sim0.57~C}$, similar to other \lithiumCrystalMath~crystals~\cite{Gillot2022}. The theoretical period calculated with measured values for the refractive indices of \lithiumCrystalMath \cite{Shen1992}, using a crystal length of 40 mm, a temperature at 20 C, and a laser wavelength of 1.0795 $\mathrm{\mu m}$, is 0.54 C. A bias tee enables application of a DC voltage along with the radio frequency drive at 4 MHz. Bandwidths up to 100 kHz are possible but not needed in the servo design.

\begin{figure}
    \centering
    \includegraphics{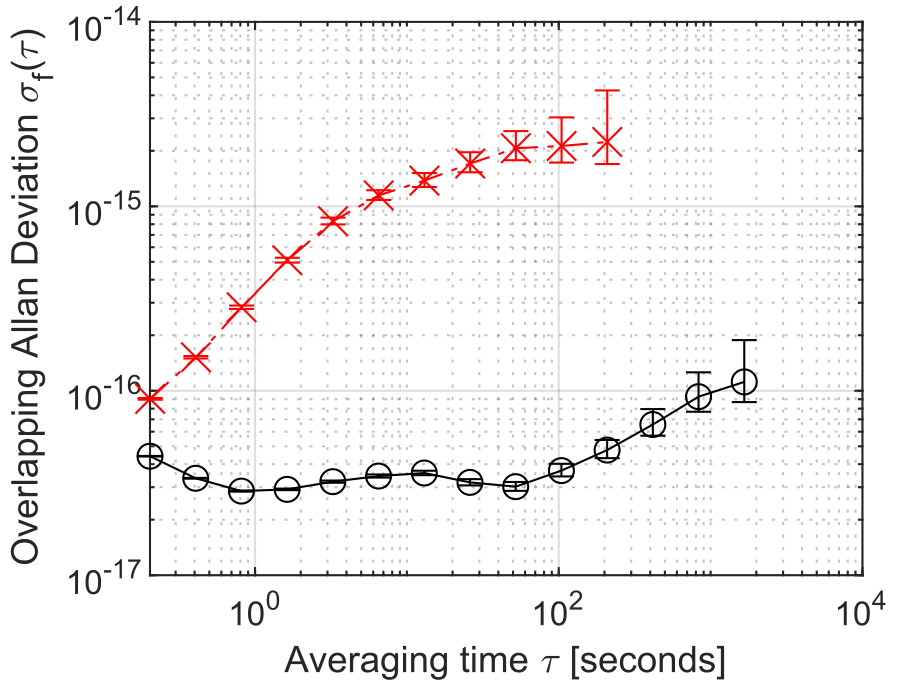}
    \caption{Residual amplitude modulation (RAM) servo performance converted to an Allan deviation of the cavity stabilized laser. Measurements with RAM servo off (red x markers) and with the RAM servo on (black circles) are taken using the PDH detector as an out-of-loop sensor for the RAM when the laser is tuned off resonance from the optical cavity. The on-resonance PDH error signal slope converts the measured offset voltage to frequency deviations. The servo suppresses the effect of drifting RAM to a fractional frequency instability of $\mathrm{3\times10^{-17}}$ up to the 100 s timescale. }
    \label{fig:ramOnOFf}
\end{figure}

By choosing a temperature that maximizes the peak-to-peak variation of RAM as a function of DC voltage, the transfer function of the DC voltage is first-order independent of the temperature. Then, the EOM DC voltage is servoed to minimize RAM based on the RAM monitor photodiode. Because temperature does not affect the servo gain, the servo operates without intervention for several weeks to months. Out-of-loop RAM is recorded via the PDH error signal off-resonance from the cavity resonance. The results are shown in Fig.~\ref{fig:ramOnOFf}. The PDH slope on-resonance is used to convert the the off-resonance data into a frequency fluctuation time series. The optical power is equal in both cases and stabilized to have a consistent PDH slope. Fig.~\ref{fig:ramOnOFf} shows a RAM induced stability at the $3\EE{-17}$ level with the servo on, which corresponds to 6.9 ppm fluctuations when referenced to the cavity linewidth \cite{Gillot2022}. Without the active servo, RAM induced noise would limit the cavity stability. The servo parameters were optimized for performance at the 0.1 s and 10,000 s time scales, testing various PI corners and gain values. A low 200 Hz PI corner and high gain gave the best results for these time scales.

\subsection{Vibrations}

\subsubsection{Acceleration Sensitivity \label{sec:vibAcc}}

The cavity length is expected to be insensitive to accelerations in the horizontal plane ($x$ and $y$) due to rotational symmetry.  For the vertical direction, acceleration sensitivity depends on the orientation of the spacer crystalline axes relative to the mounting points and it needs to be nulled experimentally.  

\begin{figure}[ht]
\centering
\includegraphics[width=85mm]{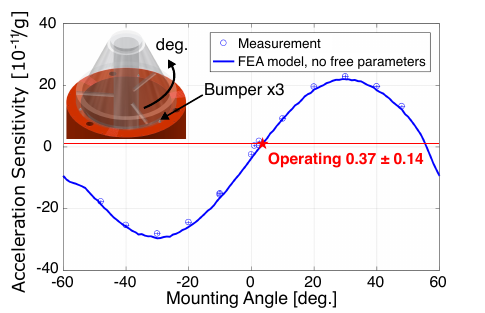}%
\caption{\label{fig:accSens} Vertical acceleration sensitivity. 
Results from finite element analysis (FEA) (blue periodic line) are compared to experimental measurements (points). The top left inset is a CAD drawing of the optical cavity on its mount with three teflon bumpers as supports. The vent holes  of the cavity and markings on the cavity mount are used to reference the angle of the cavity. The measured values closely follow the FEA model with no free parameters. The system at room temperature is operated near the red star, which has a measured acceleration sensitivity of $\mathrm{(3.7 \pm 1.4)\times10^{-12}/g}$.}%
\end{figure}

To calculate the expected vertical acceleration sensitivity, a finite element model including the cavity spacer, mirror substrates and PTFE bumpers is developed.  In the model, the following elasticity tensor for sapphire is used in Voigt notation,
\begin{gather}
 \boldsymbol{c}
 =
  \begin{bmatrix}
   c_{11} & c_{12} & c_{13} & c_{14} &  0& 0 \\
   c_{12} & c_{11} & c_{13} & -c_{14} & 0& 0\\
   c_{13} & c_{13} & c_{33} & 0       & 0& 0\\
   c_{14} & -c_{14} &  0     & c_{44} & 0& 0\\
    0& 0 & 0      & 0 & c_{44} & c_{14} \\
     0&  0&  0     & 0 & c_{14} & \scriptstyle{\frac12(c_{11}-c_{12})} \\
   \end{bmatrix},
\end{gather}
where $c_{11} = 497.5$~GPa, $c_{12} = 162.7$~GPa, $c_{13} = 115.5$~GPa, $c_{14} = 22.5$~GPa, $c_{33} = 503.3$~GPa, and $c_{44} = 147.4$~GPa~\cite{gladden2004}.We use the room temperature Young's modulus $E_{PTFE}=500$~MPa and Poisson ratio $\sigma_{PTFE}=0.41$ for PTFE in our model to match  our initial measurements which are at room temperature.  At cryogenic temperatures, PTFE is significantly stiffer but maintains some compliance~\cite{hartwig1995}. In the model, a common acceleration was applied vertically at the support surfaces and the distance between the center of the mirror substrates was numerically calculated.  Results are shown as the solid line in Fig.~\ref{fig:accSens}.    

\begin{figure}[ht]
\centering
\includegraphics[width=85mm]{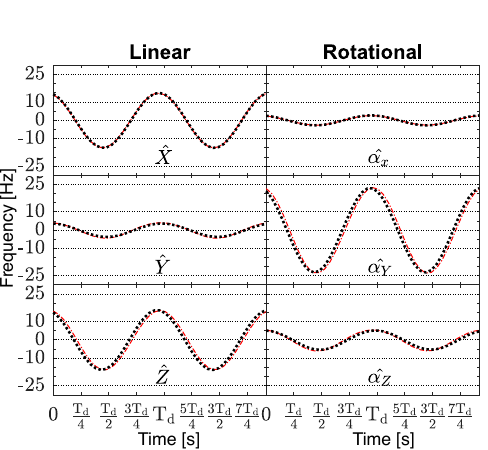}
\caption{\label{fig:accelerationDrives} Acceleration sensitivity measurement at 10~K. For each panel, a different direction of motion is driven by the AVI system at a frequency of 4 Hz ($\mathrm{T_d=1/f}$) with acceleration amplitudes near 0.01~m/s$^2$. Fine dotted lines at $\mathrm{\pm}$~10 Hz and $\mathrm{\pm}$~25 Hz are to guide the reader. The black dots show the measured frequency of the cavity-stabilized laser, and the red lines show a global fit to the data in all panels to the function $f = \sum_{i} c_i a_i$ where $i \in \{{\mathrm{\hat{X}}, \mathrm{\hat{Y}}, \mathrm{\hat{Z}}, \mathrm{\hat{\alpha}_X}, \mathrm{\hat{\alpha}_Y}, \mathrm{\hat{\alpha}_Z}}\}$.  Here, $f$ is the measured frequency, $c_i$ is the fit parameter representing the acceleration sensitivity coefficient for accelerations in the $i$ direction, and $a_i$ is the measured acceleration in the $i$ direction.  We observe a small residual phase shift between fits and the frequency data $\mathrm{\hat{\alpha}_Y}$ and $\mathrm{\hat{Z}}$, which is currently unexplained.}
\end{figure}

Experimentally, acceleration sensitivity values are determined by simultaneous measurements of six accelerometers and the cavity frequency.  The cavity is accelerated by applying a sinusoidal drive (4 Hz - 15 Hz) to the active vibration isolation (AVI) legs along one of the linear or rotational degrees-of-freedom. The acceleration data are transformed to components of rigid-body motion in a reference frame centered on the cavity center of mass. The linear and rotational accelerations and frequency excursions are measured as a function of mounting angle. Example data for one mounting angle is shown in Fig.~\ref{fig:accelerationDrives}. Because the AVI legs have crosstalk between the different degrees of freedom when driven, we perform a simultaneous fit of all the acceleration sensitives to the complete data set that includes driving each of the degrees of freedom. For each point in Fig.~\ref{fig:accSens}, x-direction error-bars stem from uncertainties in the measurement of the angle in the apparatus while y-direction error-bars are calculated from the uncertainty in acceleration noise. The lowest measured z-direction acceleration sensitivity value is $\mathrm{(3.7\pm1.4)\times10^{-12}/g}$. 

To measure the cavity response to vertical accelerations as a function of angle, initial measurements were made at room temperature and atmospheric pressure to allow for multiple points to be measured in a day, without a need for pumping down a vacuum chamber or cooling a cryostat.  These were performed with 15 Hz drive frequency and accelerations near 0.01 m/s$^2$.  The results are shown in Fig.~\ref{fig:accSens}, where good agreement with the FEM is observed. The orientation of the cavity in the cryostat was adjusted based on these measurements to null the vertical acceleration sensitivity (Fig.~\ref{fig:accSens}, red star).

Final measurements of the acceleration sensitivity were taken after cooling the cavity to 10 K.  Multiple data sets were taken to determine the mean and standard deviation of the acceleration sensitivity for each direction.  An example data set is shown in Fig.~\ref{fig:accelerationDrives}, and the acceleration sensitivity values are presented in Table~\ref{tab:accSens}.  There are slight changes in sensitivity values when cooling from 300 K down to 10 K, which might be a result of the difference in the stiffness of the PTFE bumpers.

\begin{table}
\centering
\begin{tabular}{ | >{\centering\arraybackslash}p{2cm} | >{\centering\arraybackslash}p{2cm} | >{\centering\arraybackslash}p{2cm}| }
\hline
Direction & Magnitude & Units\\
\hline
$\mathrm{\hat X}$           &  $\mathrm{3.51 \pm 0.41}$         & $10^{-11}$/g\\
\hline
$\mathrm{\hat Y}$           &  $\mathrm{-0.12 \pm 0.34}$        & $10^{-11}$/g\\
\hline
$\mathrm{\hat Z}$           &  $\mathrm{-0.75 \pm 0.10}$        & $10^{-11}$/g\\
\hline
$\mathrm{\hat \omega_x}$    & $\mathrm{-4.5 \pm 1.8}$       & $10^{-11} \mathrm{s^2/rad}$\\
\hline
$\mathrm{\hat \omega_y}$    &  $\mathrm{37.5 \pm 3.6}$      & $10^{-11} \mathrm{s^2/rad}$\\
\hline 
$\mathrm{\hat \omega_z}$    & $\mathrm{0.2 \pm 3.0}$        & $10^{-11} \mathrm{s^2/rad}$\\
\hline
\end{tabular}
\caption{\label{tab:accSens} Measured acceleration sensitivity coefficients at 10~K. The uncertainties are the standard deviation of multiple measurements.}%
\end{table}

\subsubsection{Vibration Isolation}
Because the measured acceleration sensitivity is at the $\mathrm{10^{-11}}/g$ scale, to reach the thermal noise limit, the cavity must be isolated from vibrations at the $\mathrm{\sim100~ng}$ level. In the current system, transmitted vibrations are minimized by careful alignment of the pulse-tube in the helium exchange space and are minimized by using active vibration isolation legs. The optical table is surrounded in an acrylic enclosure with foam padding to dampen acoustic noise. The cryocooler and stand also have an enclosure to separate the pulse-tube acoustic noise. During operation, the pulse-tube is rigidly mounted to a stand that is bolted to the floor, so vibrations travel primarily through the floor. The optical table rests on active vibration isolation (AVI) legs, which damp transmitted vibrations from the ground for frequencies above 1 Hz. A commercial feedforward system is used to reduce transmitted vibrations from the ground.

\begin{figure}[htbp]
    \includegraphics[]{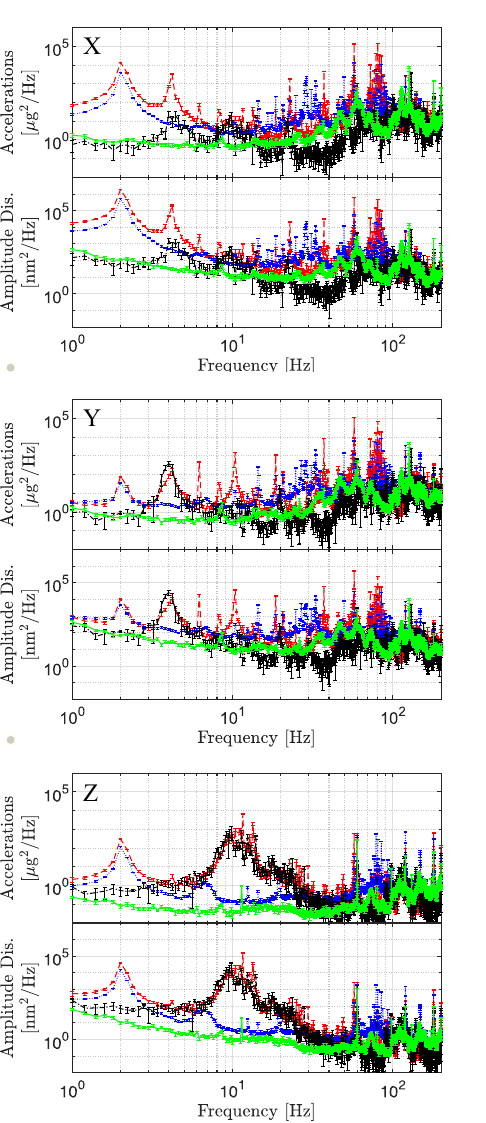}

   \caption{\label{fig:psdAccel}Measured power spectral density (PSD) of linear environmental accelerations in the $\mathrm{\hat{X}}$ (top panel), $\mathrm{\hat{Y}}$ (middle panel), and $\mathrm{\hat{Z}}$ (lower panel) directions. Black points connected by dot-dashed lines are measured with both the AVI legs and cryostat are off, green points connected by solid lines are measured with the AVI legs are on and the cryostat is off, red points connected by dashed lines are measured with the AVI legs are off and the cryostat is on, and blue points connected by dotted lines are measured with both the cryostat and AVI legs are on. Measurements are displayed in both acceleration and displacement units.}

\end{figure}

Measurements of $\mathrm{\hat{X}}$-accelerations, $\mathrm{\hat{Y}}$-accelerations, and $\mathrm{\hat{Z}}$-accelerations are shown in Fig.~\ref{fig:psdAccel}. Separate curves display measurements performed for each combination of the AVI legs and cryocooler being off and on. The power spectral density data is shown both for the directly measured accelerations and the calculated amplitude displacements, to compare with other experiments where reporting displacements is common. When the cryocooler is on, there are large amplitude vibrations at 2 Hz regardless of whether the AVI legs are on. This frequency is associated with the pulse-tube mechanical vibration. The AVI legs reduce noise above 10 Hz, but at 40 Hz the noise is higher than if the AVI legs were off. Effects beyond 200 Hz are similar in all cases, indicating both that the cryocooler does not introduce significant vibrations and the AVI legs do not significantly suppress vibrations at these frequencies. There are appreciable $\mathrm{2^{nd}}$ harmonics at 4 Hz in the $\mathrm{\hat{X}}$- and $\mathrm{\hat{Y}}$-directions. Mechanical resonances above 4 Hz are significantly reduced when the AVI legs are turned on.

\subsection{Temperature Fluctuations}

Temperature fluctuations of the cavity couple to frequency noise through the cavity coefficient of thermal expansion (CTE), which was measured to be $7.8(2)\times10^{-10}$ at the operating temperature of 10.5 K~\cite{valencia2023}.  The thermal design uses active stabilization and passive heat shields to minimize temperature fluctuations of the optical cavity. There are four heat shields. The inner two shields are the cavity mount and the inner vacuum chamber, which passively isolate the cavity.  The outer two shields, heat shield and cold plate, are temperature stabilized at 160~K and 10.5~K with residual, out-of-loop temperature fluctuations of approximately 1~mK and 100 $\mathrm{\mu}$K, respectively. An acrylic enclosure that surrounds the optical table is temperature stabilized to the 10~mK level with thermo-electric coolers.

The cryostat is designed to minimize convective, radiative, and conductive heat-transfer between heat shields. To mitigate convective heat transfer, the outer vacuum chamber is pumped down to less than $\mathrm{3\times10^{-7}~mbar}$.  Radiative heat transfer is minimized by wrapping the outermost heat shield in mylar, and polishing the surfaces of all shields. Optical BK-7 windows in each of the outer heat shields and the inner vacuum chamber attenuate room temperature blackbody radiation from heating the optical cavity. To balance the conductive heat transfer necessary for cooling down and the long thermal time constants for suppressing temperature fluctuations, the posts supporting the inner vacuum chamber and the cavity mount are made of stainless steel.  Indium foil is placed between material interfaces to maximize thermal contact. 

\begin{figure}[htbp!]
\centering
\includegraphics[width=85mm]{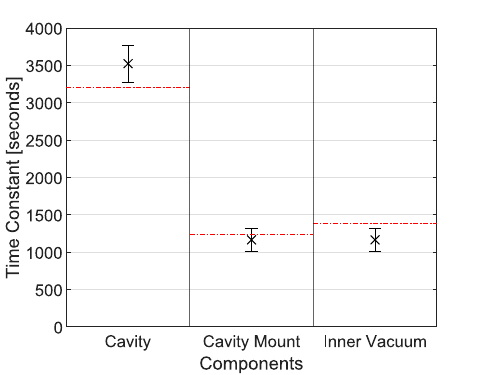}%
\caption{\label{fig:timeconstants} Thermal time constants. The dotted red lines are the calculated values. The black crosses are the measured values with error bars that represent 1-$\sigma$ fit uncertainties. Thermal contact resistances are not included in this model, but they would tend to make the calculated values larger than the experimental values and are not found to be significant here.}
\end{figure}

Each passive heat shield filters the temperature fluctuations that are transmitted through it.  Quantitatively, this filter can be modeled as cascaded, RC, electrical, low-pass filters, where the order of the circuit is determined by the number of heat shields plus one for the cavity itself. Resistances correspond to the inverse of the thermal conductance between the shields, and capacitances correspond to the heat capacities of the shields. This circuit model can be used to calculate the thermal time constants of the system. Experimentally, the time constants are measured by applying a temperature step to the innermost temperature-stabilized shield, measuring the cavity resonance frequency as a function of time, and fitting the results to the circuit model. A partial fraction expansion isolates each time constant to be used as a fitting parameter~\cite{valencia2023}. 
Fig.~\ref{fig:timeconstants} shows the calculated and measured time constants for the cavity, cavity mount, and inner vacuum chamber.  The measured values are 3500(400) s, 1350(100) s, 1350(100) s, respectively. Because of the degenerate time constants, the partial fraction expansion was also modified to account for degeneracy and was re-fitted but there was no impact on the obtained time constants. Error bars are determined from fitting uncertainties of the partial fraction expansion. Subtracting the fits with the measured values showed residuals within 5\% of the measured frequency response. The red-dashed lines display theoretical time constants.

\section{Cavity stability measurements}
The cryogenic cavity frequency stability is measured by comparing it with two room-temperature cavities using the three-cornered-hat (TCH) method \cite{gray1974}.  The underlying assumption is that the noise of the three oscillators is uncorrelated, so that Allan variance of the frequency difference of two oscillators is the sum of the Allan variances of the individual oscillators, $\mathrm{\sigma_{a,b}^2=\sigma_a^2+\sigma_b^2}$. A linear combination of three frequency-difference Allan variances can be used to isolate each individual oscillator stability.  For stability measurements beyond an averaging time of 10 s, a comparison with atom-stabilized light is made from a Yb-lattice clock~\cite{mcgrew2018}, which exhibits a stability of $1.5\times10^{-16}/\sqrt{\tau/\rm{s}}$.

\begin{figure}[ht]
\centering
\includegraphics[width=85mm]{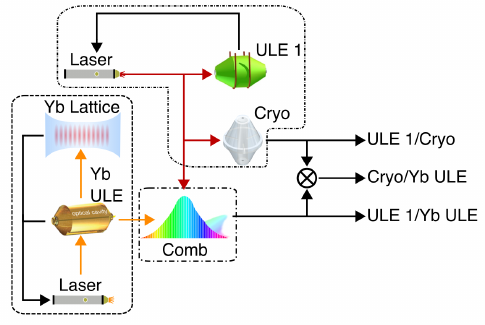}%
\caption{\label{fig:opticalBeatnotes} Setup for measurements of the cryogenic cavity's frequency stability.  The stability is measured with respect to a room temperature cavity at 1069~nm labeled ULE 1 and a room temperature cavity at 1156~nm labeled Yb ULE.  For measurements against the optical clock, the Yb ULE cavity is stabilized to a Yb optical lattice clock.  Fluctuations of the frequency difference between ULE 1 and the cryogenic cavity are obtained from the frequency of the second AOM in Fig.~\ref{fig:oscarPDHsetup}, which is used to lock the light that is prestabilized to ULE 1 to the cryogenic cavity. Fluctuations of the frequency difference between ULE 1 and Yb ULE are obtained from the frequency of a beat note between the laser stabilized to ULE 1 and a frequency comb stabilized to the Yb ULE cavity light.  These two RF frequencies are measured using dead-time-free counters, and fluctuations of the frequency difference between the cryogenic cavity and Yb ULE are obtained by subtracting the two measured frequencies in post processing.  Different components are located in three different rooms in the same building, which are denoted by dashed lines.}
\end{figure}

Fig.~\ref{fig:opticalBeatnotes} shows the setup for these measurements.  The cryogenic cavity (Cryo) and a room-temperature cavity (ULE 1) are located in one room.  A 1069~nm fiber laser is locked to ULE 1\cite{young1999} and a portion of the pre-stabilized laser light is picked off and locked to Cryo using the second AOM in Fig.~\ref{fig:oscarPDHsetup}.  A third cavity (Yb ULE) is used to stabilize an 1156~nm amplified extended-cavity diode laser (ECDL) located in a separate room.  Both the 1069~nm laser and 1156~nm are delivered to an optical frequency comb~\cite{nardelli2023} via stabilized optical fiber paths~\cite{ma1994}.  By stabilizing this comb to one laser and measuring the beatnote frequency with the second laser, an RF signal with the relative frequency noise between ULE 1 and Yb ULE is generated. For measurements of the long-term drift of the cryogenic cavity, Yb ULE is further stabilized to a Yb optical lattice clock \cite{mcgrew2018}. The third beatnote, between Yb ULE and Cryo, is generated by frequency mixing the first two in post-processing as depicted in Fig.~\ref{fig:opticalBeatnotes}. The three rooms where the lasers are located are in the same building, but each room has independent concrete slab foundations and independent air conditioning systems so environmental vibrations and temperature fluctuations are largely uncorrelated between them.

\begin{figure}[ht]
\centering
\includegraphics[width=85mm]{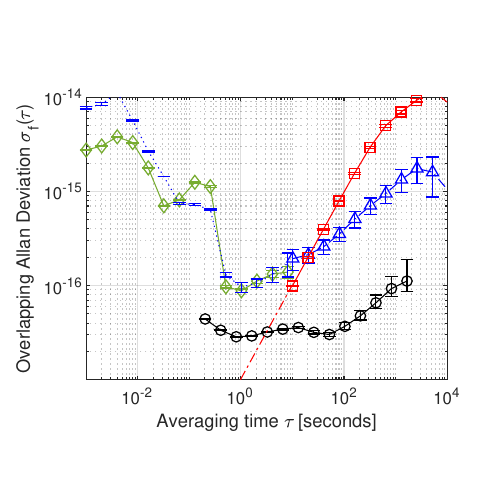}%
\caption{\label{fig:shortTermResults} Measured cryogenic cavity frequency stability. Blue points at averaging times below 10~s are TCH measurements of the cryogenic cavity frequency stability.  Blue triangles at averaging times above 10~s are frequency difference measurements between the cryogenic cavity and the Yb lattice clock stabilized cavity.  Green diamonds are the estimated contribution of accelerations to the cryogenic cavity frequency instability.  Black circles are the estimated contribution of RAM to the cryogenic cavity frequency instability.  Red squares are the estimated contribution of temperature fluctuations to the cryogenic cavity frequency instability. The red-dashed line is an extrapolation of the $\tau$ dependence of the estimated temperature fluctuations. There is 1st-order drift removal for the estimated acceleration-induced noise and the TCH measurements of the cryogenic cavity stability.}
\end{figure}

The measured stability of the cryogenic cavity is shown as the blue markers in Fig.~\ref{fig:shortTermResults}.  Also shown are the estimated contributions of RAM, accelerations, and temperature fluctuations to the cavity instability.  RAM measurements are converted to frequency using the measured PDH error signal slope; components of acceleration are converted to frequency using the measured acceleration sensitivity coefficients; and temperature measurements are converted to frequency using the measured time constants and cavity CTE.  At averaging times below 10~s the cryogenic cavity instability is primarily limited by accelerations, while at longer averaging times it is primarily limited by temperature fluctuations.  At 1~s averaging time, the fractional frequency instability is $9.6(1.2) \times 10^{-17} \approx 1\times 10^{-16}$.

At long averaging times temperature fluctuations should limit the cryogenic cavity, but the observed stability is better than the prediction based on temperature measurements. Pearson-product correlations between lab sensors, like accelerations, various temperature sensors, and servo outputs, and measured frequency data were made, but only weak correlation products < |0.2| were found and were not suggestive. To explain the high performance at 1000 s based on thermal filtering, one of the thermal time constants would need to be approximately 10 times higher than what was measured.  This is unlikely based on the close correspondence between the measured and calculated time constants (Fig.~\ref{fig:timeconstants}).  An error in the measurement of the cavity CTE is similarly unlikely.  However, an error in the measurement of temperature changes that drive the thermal expansion cannot be ruled out because of the difficulty of accurately measuring temperature changes at the level below 1 mK represented here. Because the frequency of the cavity resonance can be measured precisely when compared to an atomic clock, this is most reliable estimate we have of the cavity temperature changes.  The stability of $1\times10^{-15}$ at a 1000 s averaging time corresponds to a bulk temperature stability at the cavity of 1.4~${\rm \mu K}$.

\section{Conclusion}

Here, the design and initial characterization of a cryogenic sapphire optical cavity, with an expected thermal noise limit below $\mathrm{10^{-17}}$ is presented. The main technical noise sources including RAM, accelerations, and temperature fluctuations, and the sensitivity of the cavity to these sources of noise were characterized. This work demonstrates, for the first time, control of all noise sources in a cryogenic sapphire optical cavity at a level sufficient to achieve instability of $1\times10^{-16}$ ~\cite{He2023}. Further improvements in temperature control, vibration isolation or feedforward \cite{leibrandt2011}, and RAM will be needed to reach the thermal noise limit. By understanding the noise in the temperature sensors and increasing the mass of the optical breadboard to mitigate accelerations, there are possibilities to reach the thermal noise floor of this system. 

Recent reports \cite{yu2022} have found length fluctuations in cavities based on GaAs/AlGaAs mirrors that are larger than the predicted thermal noise level and at a level that may limit the performance of the cavity described here.  If this is the case it will be possible to circumvent much of their effects using alternative laser-locking methods \cite{kedar2022}.


\begin{acknowledgments}
This work is supported by the National Institute of Standards and Technology and the Office of Naval Research  (Grant Number N00014-20-1-2513).  We acknowledge experimental support and useful discussions from the groups of A. Ludlow and T. Fortier, and we thank L. Sonderhouse and T. Grogan for feedback on the manuscript. The data that support the findings of this study are available from the corresponding authors upon reasonable request. 
\end{acknowledgments}

\bibliography{aipsamp.bib}

\begin{thebibliography}{55}
\bibitem [{A Ultra Low Vibration cryostat was designed by ColdEdge
  Technologies. The inner modules were designed by the authors.()}]{coldedge}%
  \BibitemOpen
  A Ultra Low Vibration cryostat was designed by ColdEdge Technologies. The
  inner modules were designed by the authors.,\ \href@noop {} {}\bibinfo {note}
  {Certain commercial equipment, instruments, or materials, commercial or
  non-commercial, are identified in this paper in order to specify the
  experimental procedure adequately. Such identification does not imply
  recommendation or endorsement of any product or service by NIST, nor does it
  imply that the materials or equipment identified are necessarily the best
  available for the purpose.}\BibitemShut {Stop}%
\end{thebibliography}

\end{document}